\documentclass[aps,prb,twocolumn,floatfix,showpacs]{revtex4-1}
\usepackage{graphicx,color,amsmath,amssymb}
\usepackage{hyperref}
\begin{document}
\title{Incompressible States of the Interacting Composite Fermions in Negative Effective Magnetic Fields at $\nu=\frac{4}{13},\,
 \frac{5}{17},\, \text{and}\, \frac{3}{10}$}
\author{Sutirtha Mukherjee and Sudhansu S. Mandal}
\affiliation{Department of Theoretical Physics, Indian Association for the Cultivation of Science, Jadavpur, Kolkata 700 032, India.}
\date{\today}
\begin{abstract}
By developing an algorithm for evaluating the basis states for the composite fermions with negative effective magnetic field, we perform the 
composite-fermion-diagonalization study for the fully spin-polarized fractional quantum Hall states at the filling
factors $\nu = 3/10$, 4/13, and 5/17 in the range $2/7 <\nu < 1/3$. These observed states correspond to partially filled second effective Landau level, for the composite
fermions carrying four vortices, with filling factor $\bar{\nu} = 1/2$, 1/3, and 2/3 respectively, analogous  to the previously studied states of composite fermions with two 
attached vortices in the range $1/3 <\nu <2/5$.
We show that the character of these states in the range $2/7 <\nu < 1/3$ replicates the same in the range $1/3 <\nu <2/5$ having identical $\bar{\nu}$: 
Chiral p-wave pairing with anti-Pfaffian correlation of composite fermions carrying six quantized vortices produces incompressible state at $\nu = 3/10$;
an unconventional interaction between composite fermions, resulting from the suppression of fermion pairs with relative angular momentum three and producing 
fractional quantum Hall effect of composite fermions in the second effective Landau level with $\bar{\nu} =1/3$ and its particle-hole conjugate filling factor 2/3,
reproduces incompressible states at $4/13$ and $5/17$ filling factors. We further estimate the thermodynamic limit of the ground state energies and calculate the lowest
energy gap for neutral collective excitations of these states.

\end{abstract}

\maketitle

\section{Introduction}

The fractional quantum Hall effect (FQHE) is a fascinating physical phenomenon observed \cite{Tsui82} in high quality two dimensional
electron systems due to their highly degenerate quantized band structure in a very strong perpendicular magnetic field. 
The FQHE  occurs in a partially filled Landau level (LL) where
Coulomb interaction takes the most important part in describing \cite{Laughlin83} its physics. The FQHE at most of the filling factors in the lowest LL,
is very successfully described \cite{Jain89} as an integer quantum Hall effect (IQHE) \cite{Klitz80}
of composite fermions (CFs)
which are fermionic bound states of electrons and even ($2p$) number of quantized vortices, denoted by ${}^{2p}$CFs.
Due to the Berry phase \cite{Berry} produced by attached vortices, CFs experience an effective magnetic field
$B^* = B-2p\phi_0 \rho$, where $\phi_0$ is the magnetic flux quantum and $\rho$ is the electron
density of the system subjected to an external magnetic field $B$, and occupy some of the effective Landau levels, namely 
lambda levels ($\Lambda$Ls), formed by $B^*$. A complete filling of $n$ such $\Lambda$Ls by non-interacting CFs successfully 
describes the FQHE states observed at the filling factors $\nu = n/{(2pn\pm 1)}$ as the IQHE of CFs \cite{Jain89}. Possibilities of FQHE of CFs in a
partially filled $\Lambda$L arise when the interaction between CFs is taken into account.

The observation \cite{Pan03,Bellani10,Pan14,Cathy14} of FQHE at certain filling factors in the range $1/3 <\nu < 2/5$ suggests the role of 
the interaction between $^2$CFs in the partially filled second $\Lambda$L. Subsequently, the mechanisms \cite{Mukherjee12, Mukherjee13, Mukherjee14}
responsible for occurring FQHE of
these $^2$CFs at certain filling factors and thereby producing incompressible states for the observed filling factors have been suggested.
Any filling factor within this range may be described by the effective filling factor $\nu^*$ of $^2$CFs, as
$\nu = {\nu^*}/{(2\nu^*+1)}$,
with $\nu^{*} = 1 + \bar{\nu}$ ($0< \bar{\nu} < 1$) representing partially filled second $\Lambda$L with filling factor $\bar{\nu}$ on top of the completely
filled lowest  $\Lambda$L. 
Mukherjee {\it et al.} \cite{Mukherjee12,Mukherjee14} have shown that incompressible FQHE states with 
full and partial spin polarizations can arise at $\nu =$ 3/8, when the $^2$CFs, at $\bar{\nu} = 1/2$, capture two more vortices to turn 
into $^4$CFs and form a chiral p-wave anti-Pfaffian pairing. Recall that anti-Pfaffian pairing correlation \cite{Levin07,Lee07,Bishara09} is degenerate with 
its particle-hole conjugate partner, namely Pfaffian pairing, \cite{Moore91} for any two-body interaction. Therefore, three or higher body interaction of $^2$CFs
in the half filled $\Lambda$L are responsible for breaking the degeneracy.  Whereas an unusual interaction between the $^2$CFs, supporting W\'ojs,
Yi, and Quinn (WYQ)\cite{Wojs04} mechanism of an unconventional suppression of CF pairs with relative angular momentum three,
at $\bar \nu =$ 1/3 and 2/3, creates \cite{Mukherjee13} fully polarized incompressible states at $\nu =$ 4/11 and 5/13, respectively.

Minima in the longitudinal resistance, {\it albeit} lack of the signature of Hall plateaus, observed by Pan.
{\it et al.}\cite{Pan03} a decade ago, have also indicated the existence of very weak FQHE states at 3/10, 4/13 and
5/17 filling factors in the lowest LL. Later, a resonant inelastic light scattering experiment\cite{West06} has 
supported the emergence of higher order FQHE states in the corresponding filling range $2/7 <\nu <1/3$.
These filling factors are described in terms of partially filled $\Lambda$Ls produced by negative effective magnetic field experienced by $^4$CFs
as $\nu = \frac{\nu^*}{4\nu^*-1}$ with $\nu^* = 1+\bar{\nu}$. The FQHE at $\nu$ = 3/10, 4/13, and 5/17 may be described by the FQHE of
$^4$CFs with a negative effective magnetic field at $\bar{\nu}$ = 1/2, 1/3, and 2/3 respectively, {\it a la}, FQHE of $^2$CFs with 
positive effective magnetic field at $\bar{\nu}$ = 1/2, 1/3, and 2/3 describes \cite{Mukherjee12,Mukherjee13} FQHE
at $\nu$ = 3/8, 4/11, and 5/13. One may thus expect that the mechanisms responsible for FQHE in the range $1/3 < \nu <2/5$ will be
the same for FQHE in the range $ 2/7 < \nu < 1/3$ having same $\bar{\nu}$.
This expectation is however not yet theoretically met, primarily because of the technical difficulties regarding FQHE states with negative 
effective magnetic field for $^4$CFs.
In this paper, we have examined the possibility of forming incompressible FQHE states along with their mechanisms at these filling factors.


We have modified previous algorithm \cite{Simon12} for determining the projected basis states to an extent that we remove the issue of precision relating to very small
values of certain variables. This has helped us to 
construct the fully spin polarized low energy CF basis states at the lowest LL filling factors $\nu$ = 3/10, 4/13, and 5/17,
where each of the electrons captures four quantized vortices to turn into a ${}^{4}$CF and experiences a negative effective magnetic field
which in turn produces   $\nu^{*} = 3/2, 4/3$ and $5/3$ respectively. Then, by the method of diagonalization of
full Coulomb Hamiltonian in these truncated low energy CF basis, known as composite fermion diagonalization (CFD)\cite{Mandal02}, 
we show that the incompressible FQHE states should occur at all these three filling factors with appropriate correlations between CFs 
in their second $\Lambda$L.
We further estimate the interaction energy of the ground states in the thermodynamic limit, and minimum energy
gaps for neutral excitations from the CFD spectra.

The rest of the paper is organized as follows. In Sec. II, we show the general construction of the relationship between number of flux quanta 
and number of particles for fully spin polarized states
of ${^4}$CFs experiencing 
a negative $B^*$ in the range of filling factors $ 2/7 < \nu < 1/3 $. The CFD technique \cite{Mandal02} used for the calculation of low-lying spectra is reviewed
in Sec. III. Origin of the appropriate correlations and occurrence of incompressibility at $\nu = 3/10$ are described
in Sec. IV, while Sec. V contains the similar descriptions for $\nu = 4/13$ and $5/17$. We then estimate the thermodynamic limit of the ground state 
energy and minimum energy gap for neutral excitation at these filling factors in Sec. VI and summarize our results in Sec. VII.
Appendix A is devoted for reviewing Jain-Kamilla \cite{JainKamilla97} and Davenport-Simon \cite{Simon12} algorithms and the description of a modified algorithm developed by us for studying FQHE
states with negative effective magnetic fields. In Appendix B, we estimate thermodynamic limit of the interaction energy for the previously proposed
parafermionic wave function \cite{Read_Rezayi,Cappelli,Jolicoeur} at $\nu = 5/17$ and find that this energy is reasonably higher than our calculated 
energy using CFD technique.

\section{Flux-particle relations} 

When $B < 4\phi_0\rho$, the effective magnetic field for the $^4$CFs becomes negative, {\it i.e.}, $B^* <0$. In that case, the electronic filling factors
are related with the CF filling factors $\nu^*$ as
\begin{equation}
\nu = \frac{\nu^{*}}{4\nu^{*} - 1}.
\label{nuJ4negB}
\end{equation}
When $\nu^{*} = 1 + \bar{\nu}$ with $0< \bar{\nu} <1$ being the filling of the second $\Lambda$L,
we get the desired filling factor in the range  $ 2/7 < \nu < 1/3 $. A complete filling of the lowest $\Lambda$L along with a partial 
occupation of second $\Lambda$L at $\bar{\nu}$ = 1/2, 1/3, and 2/3 form fully spin polarized states at ${\nu}$ = 3/10, 4/13, and 5/17, 
respectively. 

In a spherical geometry\cite{Haldane83}, $N$ electrons moving on the surface
of a sphere experience a radial magnetic field generated by a magnetic monopole of strength $Q$ placed at the center of that sphere.
An integer number of flux quanta denoted by $2Q$ pass through the surface of the sphere and the ${}^{4}$CFs experience an effective
negative flux  $2q = 2Q-4(N-1)$ in units of magnetic flux quantum. The total number of single particle states in the lowest and the second $\Lambda$L are
$2\vert q\vert + 1$ and $2(\vert q \vert+1)+1$, respectively. The fractional filling $\bar{\nu}$ of the second $\Lambda$L with $\bar N$ CFs is related
to the effective flux $2q$ as
\begin{equation}
2|q|=\bar{\nu}^{-1}\bar{N} - (\lambda +2) 
\end{equation}
where the ``flux-shift'' parameter $\lambda$ determines topology of the fractional quantum Hall state of $^4$CFs at the filling factor $\bar{\nu}$.
The total number of particles is then $N = \bar{N}+(2\vert q\vert +1)$. The flux-particle relation for the filling factor $\nu = \lim_{N\to\infty}
\frac{N}{2Q}$ in the range $2/7 < \nu < 1/3$ becomes
\begin{equation}
2Q = \nu^{-1}(N -1)-(3-\nu^{-1})(\lambda + 2).
\label{flux-particle}
\end{equation}
We will consider different sets of $(N,2Q)$ belonging to several finite size systems at certain filling factors, namely, 3/10, 4/13, and 5/17 with specific 
values of $\lambda$.


\section{Diagonalization in Low Energy CF Basis}

Exact diagonalization of the Coulomb Hamiltonian is the most accurate way to study the finite size systems, but the exponential
increase of Hilbert space dimension restricts us to use this technique for higher number of particles. 
We thus use the CFD technique which can deal with much larger systems as it works in a much smaller subspace.
This method has already been proven\cite{Mukherjee12,Mukherjee13} to be the next best candidate for calculating energy with considerable accuracy in
the neighboring filling factor range $ 1/3 < \nu < 2/5$.

In the CFD\cite{Mandal02}, the diagonalization is performed in a truncated low energy
composite fermion basis which ignores the $\Lambda$L mixing by neglecting all the configurations involving 
promotion of CFs into higher $\Lambda$Ls. This restriction helps us to work within a reduced subspace of exponentially smaller dimension than that of the
Hilbert space of the lowest LL.  As an illustration, we tabulate exact dimension of the Hilbert space for some of the systems described by $(N,2Q)$ at
the filling factors 3/10, 4/13, and 5/17 in the Table \ref{Exactdmnsn} at various angular momenta, and the reduced dimensions of these
systems in the Table \ref{CFDdmnsn}.

We first perform exact diagonalization for a system of particles and fluxes $(\bar{N},2\vert q \vert +2)$ in the lowest LL to know
the combinations of the basis states, which provide definite angular momenta. We then construct states at $\nu^*$, by elevating these basis states into
the second LL on top of the fully filled lowest LL. The angular momenta of the latter type of states will be the same as that of the former type of states
since the lowest LL is fully filled. There will be several states denoted by $\Phi_{\nu^*}^{L,\alpha}$
($\alpha$'s labeling different states with a particular angular momentum),
corresponding to each angular momentum, $L$, as shown in Table~\ref{CFDdmnsn}. We then composite
fermionize (as $^4$CFs) these states with the Jastrow factor $J^4$ where
\begin{equation}
J = \prod_{j<k}^{N}\left(u_jv_k - v_ju_k\right),
\end{equation}
with $ u_j \equiv \cos(\theta_j/2)\exp(-i\phi_j/2) $ and $ v_j \equiv \sin(\theta_j/2)\exp(i\phi_j/2) $ being two spinor variables for
fermions with angular coordinates $\theta_j$ and $\phi_j$. Finally, we construct the state $
\Psi_{\nu}^{L, \alpha}$  as 
\begin{equation}
\Psi_{\nu}^{L, \alpha} =  P_{\rm{LLL}}J^4 \Phi^{L, \alpha}_{\nu^*} \, ,
\label{wfn_psi}
\end{equation}
by projecting onto the lowest LL represented by the operator $P_{\rm{LLL}}$.

All the $\Phi$'s in Eq.(\ref{wfn_psi}) are antisymmetric many body states, which can be obtained 
from Slater determinants consisting of monopole harmonics\cite{Wu76} of the form \cite{Simon05}
\begin{eqnarray}
Y_{|q|,l,m} (u_j, v_j) &=& 
N_{|q|,l,m}\left(-1\right)^{(|q|+l-m)}  u_j^{*(|q|+m)} 
   \nonumber \\
&\times & v_j^{*(|q|-m)}
  \sum_{s=0}^{l} (-1)^s{l \choose s}   {2|q|+l \choose |q|+m+s}      \nonumber \\
&\times & (u_j^* u_j)^s (v_j^* v_j)^{l-s}  ,
\label{Yqlm}
\end{eqnarray}
where $l$ = 0, 1, ... stands for the $\Lambda$L index, $ m = -|q|-l,
-|q|- l + 1, ...., |q|+l$ labels the degenerate states of $l$th $\Lambda$L, and the normalization coefficient $N_{|q|lm}$ is given by
\begin{equation}
N_{|q|lm}= \left(\frac{(2|q|+2l+1)}{4\pi}\frac{(|q|+l-m)!(|q|+l+m)!}{l!(2|q|+l)!}
\right) ^{1/2}\;\;.
\end{equation}
We develop an efficient algorithm, described in the appendix,
for numerical calculation of these many body states $\Psi_{\nu}^{L, \alpha}$ for relatively large finite size systems studied by CFD. 
Though the invariance of total orbital angular momentum gives us the low energy correlated
many body basis states $\Psi_{\nu}^{L,\alpha}$ at different $L$ sectors independently, the basis looses its orthogonality. We follow the 
Gram-Schmidt procedure to obtain an orthogonal basis and then generate the full Coulomb Hamiltonian matrix in this reduced basis
by Metropolis Monte Carlo technique. Performing 10 Monte Carlo runs for each system with $10^7$ Monte Carlo iterations in each run, we reduce the statistical 
uncertainty to a desired level. Finally, we diagonalize this matrix in each $L$ sector separately to get the low-lying energy eigen
spectrum as well as eigenfunctions at any given filling factor.

\begin{widetext}
\begin{table}[h]
\caption{Full lowest LL Hilbert space dimensions in various angular momenta, $L$ with its $z$-component $L_z = 0$, 
for various systems ($N,2Q $) corresponding to fully spin polarized 3/10, 4/13, and 5/17 filling factors.}
\label{Exactdmnsn}
\setlength{\tabcolsep}{0.15em}
\centering
\footnotesize 
\vspace{2.0ex}
\hspace{-0.5cm}\begin{tabular}{c|| rrrrrrrrrrr}
\hline\hline
$(N,2Q)$&$L=0$&1&2&3&4&5&6&7&8&9&10\,\\ 
\hline
\hline
$(11,34)$ & 3569 &10213  & 17291 & 23853  & 30792 & 37159 & 43877 & 49973 & 56365 & 62120 & 68116\\
$(12,37)$ & 16027 & 46127 & 77978  & 107776  & 139132 & 168244 & 198803 & 226944 & 256342 & 283246 & 311217 \\
$(12,38)$ & 20903 & 60822 & 102380  & 141933   & 182870 & 221568 & 261535 & 299029 & 337548 & 373511 & 410272 \\
$(14,45)$ & 615228 & 1831756  & 3058345  & 4267188  & 5482253 & 6675794 & 7871772 & 9042557 & 10212090 & 11352897 & 12488877 \\
$(16,51)$ & 15278595 & 45750042  & 76239996  & 106577925  & 136867465 & 166938560  & 196895856 & 226569437 & 256064361 & 285212557
& 314118882 \\\hline\hline
\end{tabular}
\end{table}
\end{widetext}

\begin{table}[h]
\caption{Dimension of the reduced basis used in CF diagonalization. 
These are equal to the exact dimension of the Hilbert subspace in the lowest LL for $\bar N$ particles at the flux $2\vert q\vert +2$.}
\label{CFDdmnsn}
\footnotesize
\begin{tabular}{c rrrrrrrrrr}
\hline\hline
$(N,2Q)$&$L=0$&1&2&3&4&5&6&7&8&9\,\\\hline\hline
$(11,34)$ \& $(12,38)$&1&0&2&1&2&1&2&1&1&0\,\\
$(12,37)$ \& $(14,45)$&2&0&2&1&3&1&3&1&2&1\,\\
$(16,51)$ &3&0&4&3&6&3&7&4&6&4\,\\
$(18,57)$ \& $(20,65)$&4&1&7&5&11&7&13&9&13&10\,\\
$(20,64)$ \& $(21,68)$&4&3&10&9&16&14&19&17&21&18\,\\
$(24,77)$ &12&10&32&30&51&48&66&61&77&70\,\\
\hline\hline
\end{tabular}
\end{table}

\section{Incompressible state at { \large{\boldmath{$\nu = \frac{3}{10}$}}} }

Half filling of the second $\Lambda$L on top of a completely filled lowest $\Lambda$L forms $\nu = \frac{3}{10}$ following 
Eq.~(\ref{nuJ4negB}). The FQHE is not observed at $\nu = 1/2$ because the $^2$CFs form a gapless Fermi sea\cite{Halperin93,Kalmayer92}.
However, the attachment of vortices may overscreen the interaction between electrons; the effective interaction between CFs becomes
attractive which initiates
instability in the Fermi sea of CFs in favor of forming their paired states, as in the case \cite{Moore91,Scarola00} of half-filled second LL \cite{Willett87}. 
There are two topologically distinct candidates to describe such pairing: a Bardeen-Cooper-Schrieffer (BCS)--like Pfaffian wave function proposed by
Moore and Read\cite{Moore91}, and its particle-hole conjugated partner, the anti-Pfaffian\cite{Levin07,Lee07,Bishara09}.
Pairing of CFs in the second $\Lambda$L has already been predicted \cite{Mukherjee12} for a possible incompressible FQHE
state at fully spin polarized lowest LL filling $3/8$. Here, we show that an incompressible state is also very likely to occur
at another fully polarized even denominator filling factor $\nu = 3/10$ due to the chiral p-wave pairing of CFs in  
the second $\Lambda$L formed by negative effective magnetic flux. We find that the incompressibility arises at 3/10 due to the
following physical mechanism: the electrons capture four vortices to turn into $^{4}$CFs which experience a negative effective magnetic field
that forms a half filled second $\Lambda$L on top of a completely filled lowest $\Lambda$L; $^{4}$CFs capture two more vortices to 
turn into $^{6}$CFs which open the gap by forming a BCS like pairing in the second $\Lambda$L; this pairing gap manifests incompressibility
at the filling factor $\nu=3/10$.

\begin{figure}[h]
\vspace{1cm}\includegraphics[width=8.5cm,angle=0]{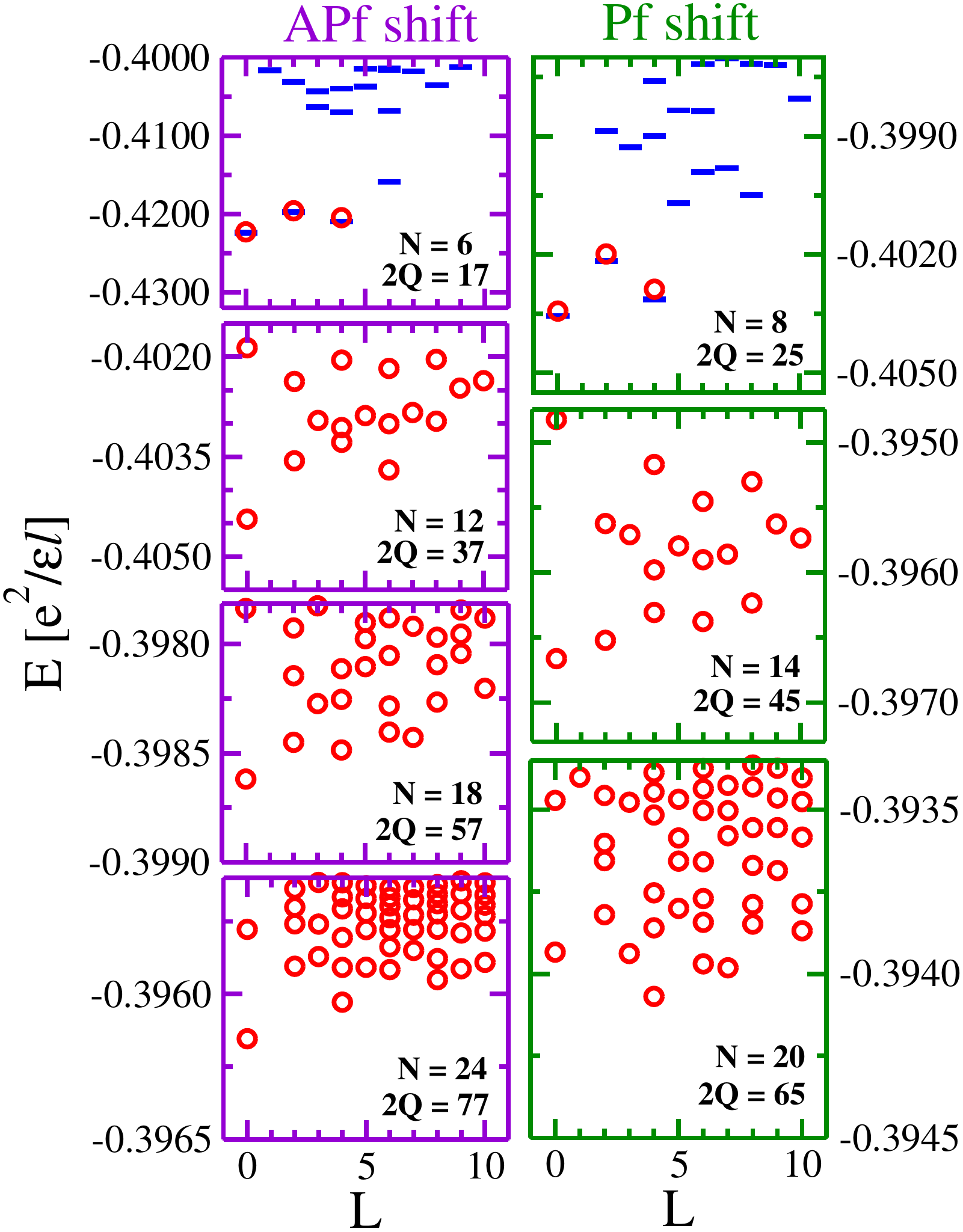}
\caption{(Color online) The low-lying energy spectra obtained by CFD with both ``Pf shift" (right panels) and ``APf shift" (left panels) at fully
spin polarized $\nu=3/10$, for systems with various values of $N$ and $2Q$ shown inside the panels. The dashes represent the
energy spectra obtained by exact diagonalization of the Coulomb interaction in the full LLL Hilbert space. The statistical uncertainty,
estimated from Metropolis Monte Carlo evaluation of integrals is less than the diameter of any circle. The energy per particle, $E$ includes the
interaction with the background.  
	}
\label{310spectra}
\end{figure}

The above pairing mechanism suggests a flux shift $\lambda$ = 3 for Pfaffian (Pf) type
and $\lambda$ = -1 for anti-Pfaffian (APf) type of pairing, referred as  ``Pf shift" and ``APf shift" respectively.
We calculate the energy eigenvalues at different angular momentum sectors for several system sizes considering both types of shifts. 
For the smallest systems ($N=6$ at APf shift and $N=8$ at Pf shift), we calculate the full energy spectra by exact diagonalization of
Coulomb Hamiltonian and compare it with the low-lying spectra obtained from CFD as shown in Fig.{\ref{310spectra}}.
The very small deviations ($\sim$ 0.05 \%) of CFD energies from exact ones indicate that CFD is a very reliable technique for obtaining
low-lying spectra at 3/10 filling. We thus calculate the  energy spectra for larger systems using CFD which allows us to calculate  
up to $N=24$ at APf shift and $N=20$ at Pf shift. 

A non-degenerate ground state at $L=0$, separated from all other states by a finite energy gap is identified \cite{JainKamilla97,Jainbook} as an incompressible state
in spherical geometry. The CFD produces a ground state at $L=0$ for all the $(N,2Q)$ values at APf shift, but not for all of them 
(see Fig.\ref{310spectra}) at Pf shift.
The low-lying energy spectra disfavor the Pf type of pairing and support APf type of pairing in the second $\Lambda$L. 
Therefore, the incompressible FQHE state at 3/10 filling occurs due to APf-type pairing correlation.

For further investigation of the possible pairing mechanism, motivated by our previous study \cite{Mukherjee12} at $\nu =3/8$, we construct
trial wave functions 
\begin{equation}
\label{wavefunction}
\Psi_{\frac{3}{10}}^{\rm trial-Pf/APf} =  P_{\rm{LLL}} J^4 \Phi^{\rm Pf/APf}_{3/2},
\end{equation}
where $\Phi^{\rm Pf/APf}_{3/2}$ represents Pf or APf wave function at $\nu^{*} = 3/2$ in a negative effective magnetic field. The Pf state in the lowest LL can be 
obtained as the zero energy ground state of a  3-body interaction Hamiltonian $V_3 = \sum P_{i<j<k}^3(3Q-3)$ with $2Q = 2N-3$, where the
three body projection operator\cite{Greiter91} $P_{i<j<k}^3(L)$ projects out any cluster of three particles $(i,j,k)$ with total
angular momentum $L$, and the APf state can be obtained by the particle-hole conjugation of Pf state. 
We construct $\Phi_{3/2}^{{\rm Pf/APf}}$ states by completely filling the lowest $\Lambda$L and elevating the Pf/APf state to the 
second $\Lambda$L.
Table {\ref{Pfovrlp}} contains the overlap of $\Psi_{\frac{3}{10}}^{\rm trial-Pf/APf}$ with the 
corresponding CFD ground state at $L=0$.
APf state has higher overlaps (excepting the smallest system) than the Pf for all the systems with same 
$q$. These higher overlaps along with the incompressibility for all the available system sizes at APf shift make a strong
case that a fully spin polarized incompressible state is possible at $\nu = 3/10$, when the composite fermions experiencing a
negative effective magnetic field form APf type of pairing in the half filled second $\Lambda$ level.

\begin{table}[h]
\caption{Overlaps between the CFD ground state $\Psi_{3/10}^{\rm{CFD}}$ and the trial wave function $\Psi_{3/10}^{\rm{trial-Pf/APf}}$
at fully spin polarized 3/10 filling. $\Psi_{3/10}^{\rm{CFD}}$ is obtained by CF diagonalization at the ``Pf shift'' and ``APf shift'';
$\Psi_{3/10}^{\rm{trial-Pf/APf}}$ is derived from the composite fermionization of Pf/APf state at fully spin polarized 3/2. 
For a special case marked by $*$ at $N$ = 20, the comparisons are given for the lowest energy state in the $L = 0$ sector as the CFD ground state occurs at $L = 4$. 
        }
\label{Pfovrlp}        
\begin{tabular}{ c  c  c  c  c  c} \hline\hline
$N$~&~~$2Q$~~&~$\bar N$~&~~$2q$~&~ $\langle\Psi_{3/10}^{\rm CFD}|\Psi_{3/10}^{\rm{trial-Pf}}\rangle$~&~
$\langle\Psi_{3/10}^{\rm CFD}|\Psi_{3/10}^{\rm{trial-APf}}\rangle $  \\ \hline\hline
6 & 17 & 2  & 3  & -- & 1.000\\
8 & 25 & 4  & 3  & 1.000 & --\\
12 & 37 & 4  & 7  & --  & 0.8585(2)\\
14 & 45 & 6  & 7  & 0.7770(2)   &--\\
18 & 57 & 6  & 11 & --  & 0.7350(3)\\
$^*$20 & 65 & 8  & 11 & 0.6890(8)   & --\\
24 & 77 & 8  & 15 & --  & 0.6509(9) \\\hline\hline
\end{tabular}
\end{table}

\section{Incompressible states at { \large{ \boldmath{ $\nu = \frac{4}{13}$}}} and \large{ \boldmath{$\frac{5}{17} $}} }


The filling factors of $^4$CFs, $\nu^* = 1+1/3$ and $1+2/3$, in a negative effective magnetic field correspond to $\nu = 4/13$ and 5/17 respectively. 
An ``unconventional" 1/3 state and its conjugate 2/3, arising from an unusual
correlation of $^2$CFs suggested by WYQ\cite{Wojs04} in the second $\Lambda$L, 
have already been found to generate incompressible FQHE states at fully spin polarized 4/11 and 5/13
filling factors\cite{Mukherjee13}. We here provide a theoretical evidence that these unconventional FQHEs of ${}^4$CFs in the second $\Lambda$L 
as well produce 
incompressible states at fully polarized 4/13 and 5/17 fillings of the lowest LL.

  \begin{figure}[h]
\vspace{1cm}\includegraphics[width=8.5cm,angle=0]{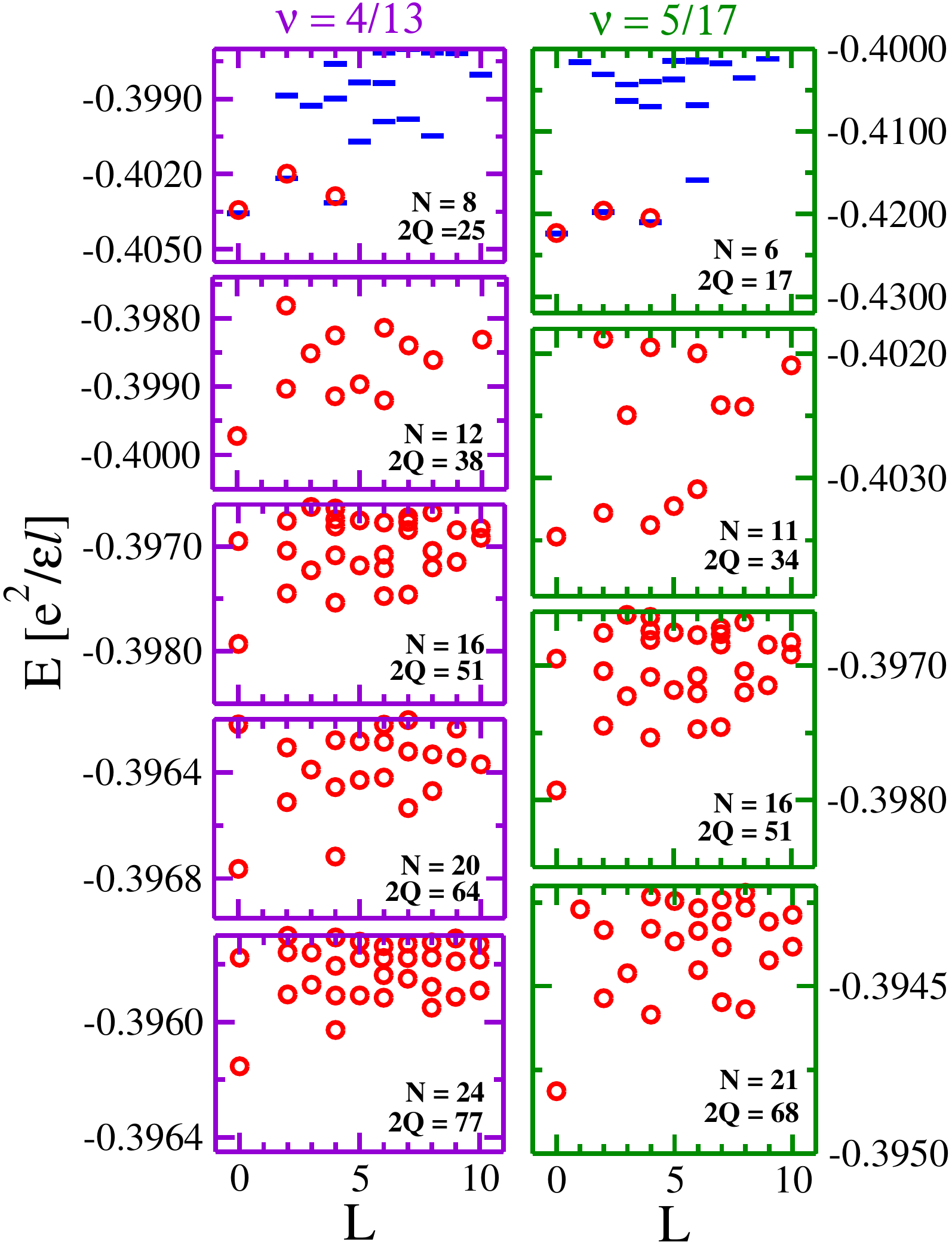}
\caption{ (Color online) The low-lying energy spectra obtained by CFD at fully spin polarized $\nu=4/13$ and 5/17 emerging out of the unconventional
FQHE of ${}^4$CFs at 1/3 and 2/3 filling in second $\Lambda$L. $N$ is the total number of electrons and $2Q$ is the total flux
quanta passing through the system. The diameter of any circle is smaller than the statistical uncertainty estimated from
Metropolis Monte Carlo evaluation of integrals. The energy per particle, $E$ includes the interaction with the background. 
}
\label{413spectra}
\end{figure}

With the flux shift $\lambda$ = 7 for unconventional 1/3 and $\lambda $ = -2 for unconventional 2/3 state, we calculate the low-lying energy
spectra at completely spin polarized $\nu$ = 4/13 and 5/17 using CFD technique and show in Fig {\ref{413spectra}}. The smallest systems ($N=8$ and 6)
for these two filling factors are same as that at 3/10 filling with Pf and APf-shift respectively. Since the
CFD spectra match (see Fig.\ref{310spectra}) quite well with the exact ones, it motivates us again to perform CFD for larger system sizes.
Figure {\ref{413spectra}} suggests  incompressible states  both at 4/13 and 5/17 filling factors as the spectra demonstrate nondegenerate ground 
states at $L=0$ with finite energy gaps for all the available system sizes. We next calculate 
the overlap of the CFD ground states for all the finite systems studied at $\nu =4/13$ and 5/17 with the corresponding trial states
\begin{eqnarray}
&&\Psi_{\frac{4}{13}}^{\rm Unconv} =  P_{\rm{LLL}} J^4 \Phi^{\rm Unconv}_{4/3} 
\end{eqnarray}
and
\begin{eqnarray}
&&\Psi_{\frac{5}{17}}^{\rm Unconv} =  P_{\rm{LLL}} J^4 \Phi^{\rm Unconv}_{5/3} 
\label{Psi_517}
\end{eqnarray}     
respectively, where $\Phi^{\rm Unconv}_{4/3}$ and $\Phi^{\rm Unconv}_{5/3}$ represent fully polarized unconventional states at 
$\nu^{*}$ = 4/3 and 5/3 respectively. The unconventional WYQ state at 1/3 filling  is the ground state of Haldane 
pseudopotential \cite{Haldane83} $V_3$ that minimizes the occupation of any two particles with relative angular momentum three,
and particle-hole 
conjugation of this state produces unconventional state at 2/3 filling. 
The overlaps are shown in table {\ref{413ovrlp}}.
Though the minimum overlap achieved is $\sim$ 81$\%$, not as large as it was in the case \cite{Mukherjee13} of fully polarized
4/11 and 5/13 states, taking together these overlaps and the nature of low-lying energy spectra, we predict that incompressible FQHE states
are highly likely to from at fully spin polarized $\nu = $ 4/13 and 5/17. These incompressible states originate from an unconventional
interaction between CFs and are well characterized by the WYQ correlation.

\begin{table}[h]
\caption{Overlaps between the CFD ground state $\Psi^{\rm {CFD}}$ and the trial wave function $\Psi^{\rm{Unconv}}$ at fully polarized
4/13 and 5/17 filling. $\Psi_{4/13}^{\rm{CFD}}$ and $\Psi_{5/17}^{\rm{CFD}}$  are obtained by CF diagonalization with unconventional
flux shifts at 1/3 and 2/3 filling of second $\Lambda$L; $\Psi_{4/13}^{\rm{Unconv}}$ and $\Psi_{5/17}^{\rm{Unconv}}$ are derived 
from the composite fermionization of unconventional states at fully spin polarized 4/3 and 5/3 filling respectively. WYQ states do not exist for $N = 4$ and 2 at 1/3 and 2/3  filling factor respectively. So, it is not possible to calculate the overlaps for the smallest systems with $N = 8$ and 6 at $\nu =$ 4/13 and 5/17 respectively. }
\label{413ovrlp}
\begin{tabular}{ c  c  c  c  c  c} \hline\hline
$N$~&~~$2Q$~~&~$\bar N$~&~~$2q$~&~ $\langle\Psi_{4/13}^{\rm CFD}|\Psi_{4/13}^{\rm{Unconv}}\rangle$~&~$\langle\Psi_{5/17}^
{\rm CFD}|\Psi_{5/17}^{\rm{Unconv}}\rangle $  \\ \hline\hline
11 & 34 & 4  & 6  & --& 1.000\\
12 & 38 & 5  & 6  & 1.000   &--\\
16 & 51 & 6  & 9 & 0.9912(1) & 0.9912(1) \\
20 & 64 & 7  & 12 & 0.9797(1) &--\\
21 & 68 & 8  & 12 & --& 0.9837(1)\\
24 & 77 & 8  & 15 & 0.8135(5) &-- \\\hline\hline
\end{tabular}
\end{table}

\section{Ground state energy and excitation gap  }

\begin{figure}[h]
\vspace{1cm}\includegraphics[width=7.5cm,angle=0]{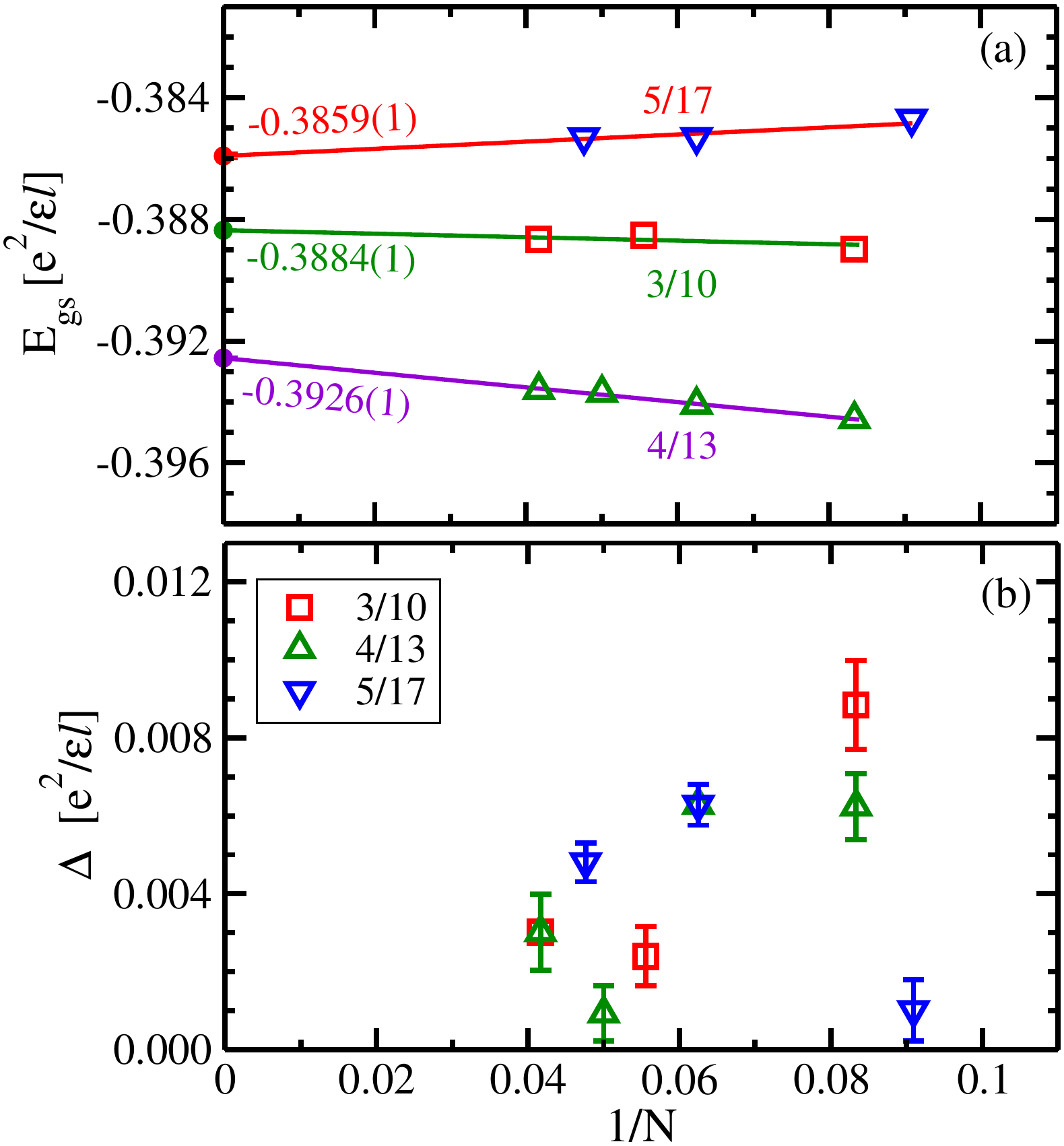}
\caption{(Color online) (a)The ground state interaction energy per particle, $\rm E_{gs}$, obtained by CFD at fully spin polarized $\nu=$ 3/10 (with APf shift), 4/13 and 5/17
are plotted against $1/N$ and extrapolated to the thermodynamic limit. (b) The minimum energy gap, $\Delta$, of neutral excitation is plotted as
function of $1/N$. We do not consider the ground state energy as well as the excitation gap of the smallest systems with N = 6 and
8 to neglect the large deviation in results due to strong finite size effect.      
}
\label{gsgap}
\end{figure}

For estimating ground state energies per electron for the incompressible FQHE states at $\nu = 4/13$, 5/17, and 3/10 in the thermodynamic limit, we first
subtract the background energy $N^2/(2\sqrt{Q})$ from the corresponding ground state interaction energies for finite systems, followed by the 
multiplication with the factor $\sqrt{2Q\nu /N}$ as the density correction and then divide by $N$ before plotting (Fig.~(\ref{gsgap}(a)) against $1/N$.
The thermodynamic limit is obtained by extrapolating the fitted lines up to $1/N \to 0$.
These energies are  $-0.3884(1)$, $-0.3926(1)$, and $-0.3859(1)$ \cite{appendix-b} in the unit of $e^2/\epsilon \ell$ (where $\epsilon$ is the background dielectric 
constant and $\ell = \sqrt{{\hbar c}/{eB}}$ is the magnetic length), at $\nu=3/10$, 4/13 and 5/17, respectively.
We also plot the lowest energy gap for neutral excitations at these filling factors in  Fig.~{\ref{gsgap}}(b). Though the results
show a consistent non-zero gap for all systems, the strong finite size effects restrict us to do a reliable extrapolation of gaps to the
thermodynamic limit. Considering the largest studied systems, we predict that the energy scale for the gap is $\sim 0.001 e^2/\epsilon \ell$,
in consistence with our recent study \cite{Mukherjee15} of neutral collective modes.
This tiny gap (about two orders of magnitude lower than the theoretically
calculated gap \cite{JainKamilla97} at neighboring 1/3 state) indicates that the interaction between the CFs in a fractionally filled $\Lambda$ level is much weaker
than that of electrons in the fractionally filled lowest Landau level.

\section{summary}   

The experimental signatures \cite{Pan03,West06}, though feeble, of the presence of FQHE states at $\nu$= 3/10, 4/13, and 5/17
within the range $2/7 < \nu < 1/3$ indicate the FQHE of $^4$CFs arising out of the interaction between them in a partially filled $\Lambda$L.
Considering a model of interacting  $^4$CFs in a negative effective magnetic
field, we have shown that an incompressible FQHE state is very likely to occur at fully spin polarized $\nu$ = 3/10, 4/13 and 5/17 due
to unconventional correlations between the composite fermions in partially filled second  $\Lambda$ level.
The character of these three states respectively replicate previously studied fully polarized 3/8, 4/11 and 5/13 FQHE states which are formed 
due to $^2$CFs in a positive effective 
magnetic field.
The ground states arise at $\nu=3/10$ due to an anti-Pfaffian type chiral p-wave pairing of CFs in their half-filled second $\Lambda$L after each of
them captures two more vortices, and at $\nu=4/13$ and 5/17 
due to WYQ type unconventional correlations between CFs in the partially filled second $\Lambda$L.
The small values of minimum energy gap for neutral excitations suggest that the states are very fragile
and thus  very high quality samples are necessary to obtain prominent quantum Hall plateaus.

 \appendix   
\section{ALGORITHMS FOR THE LOWEST LANDAU LEVEL PROJECTION}

In this appendix, we first review Jain-Kamilla \cite{JainKamilla97} and Davenport-Simon \cite{Simon12} algorithms for determining the lowest Landau level projection for the basis states of the composite
fermions when the effective magnetic field for the CFs is negative, {\it i.e.}, opposite to the applied magnetic field. We will then find that neither of these algorithms
will be suitable for finding CFD spectra for the states with $^4$CFs and negative effective field. We thus next develop an algorithm which enables us to determine the CFD spectra
for these states.

These single particle states may be obtained \cite{Simon05} from the following equation:
\begin{eqnarray}
 && Y_{\left|q\right|,l,m}(u_j,v_j){J_j}^p =  
N_{\left|q\right|,l,m}\left(-1\right)^{(q+l-m)}\frac{(2Q+1)!}{(2Q+l+1)!} \nonumber \\ 
& & \times  \sum_{s=0}^{l}(-1)^s{l \choose s}{2|q|+l \choose |q|+m+s}u_j^sv_j^{l-s} \nonumber \\
& & \times \left[\left(\frac{\partial{} }{\partial u_j}\right)^{(|q|+m+s)}\left(\frac{\partial{} }
{\partial v_j}\right)^{(|q|-m+l-s)}{J_j}^p\right]
\label{appndx_mh}
\end{eqnarray}
where $J_j = \prod_{i \ne j}^{N}\left(u_iv_j - v_iu_j\right)$ is the Jastrow factor for $j$-th particle. Because of the order of derivatives in Eq.(\ref{appndx_mh})
grows rapidly with $N$, the calculation is numerical expensive and it has many other technical problems such as apparent singularity and precision. We discuss
below the algorithms and their limitations for determining these basis states.

\subsection{Jain-Kamilla algorithm}

In Jain-Kamilla's LLL projection algorithm,\cite{JainKamilla97} the Jastrow factor is pulled through the 
derivatives and written as
\begin{eqnarray}
&&\left[\left(\frac{\partial{} }{\partial u_j}\right)^{(|q|+m+s)}\left(\frac{\partial{} } {\partial v_j}\right)^{(|q|-m+l-s)}{J_j}^p\right]
\nonumber\\
&& =  {J_j}^p\left[{ \widehat{U}_j}^{(|q|+m+s)}{\widehat{V}_j}^{(|q|-m+l-s)}\right],\nonumber
\end{eqnarray}
where 
\begin{eqnarray}
&&\widehat{U}_j = J_j^{-p}\frac{\partial{} }{\partial u_j}J_j^{p} = p\sum_k^{\prime}\frac{v_k}{u_jv_k-v_ju_k}+\frac{\partial{} }{\partial u_j}
\nonumber\\
&&\widehat{V}_j = J_j^{-p}\frac{\partial{} }{\partial v_j}J_j^{p} = p\sum_k^{\prime}\frac{-u_k}{u_jv_k-v_ju_k}+\frac{\partial{} }{\partial v_j}.
\nonumber
\end{eqnarray}
Since the exponents of $\hat{U}_j$ and $\hat{V}_j$ increase very rapidly with $N$, 
their numerical values become huge when two particles come very close during Monte Carlo steps, and thereby rejection of the subsequent moves occurs.
In particular, we can not deal with larger number of particles (e.g. $N > 18$ for 
$\nu = {2}/{7}$) using this type of projection algorithm.

\subsection{Davenport-Simon algorithm}

Davenport and Simon \cite{Simon12} developed an alternate method which avoids 
the troublesome denominators with higher exponents of $1/(u_jv_k-v_ju_k)$ when two particles come close during Monte Carlo. They begin with expanding  $J_i$ as 
\begin{eqnarray}
J_i &=& v_i^{N-1}\prod_{j \ne i}^N u_j -u_iv_i^{N-2}\sum_{j \ne i}^{N}\left(v_j\prod_{k \ne i,j}u_k\right)+\cdots \nonumber \\
   &=& \left(\prod_{j \ne i}^N u_j\right)\left[ \sum_{t=0}^{N-1}(-1)^te_{t,N-1}^i v_i^{N-1-t}u_i^t \right],
\end{eqnarray}
where $e_{t,N-1}^i$ represents a symmetric polynomial with degree $t$ in the $(N-1)$ variables of $y_j=v_j/u_j$ for $j\neq i$.
Therefore each element of a Slater determinant will have the form
\begin{eqnarray}
 & & \hat{Y}_{|q|,l,m}\left(u_i,v_i\right)J_i \propto 
\sum_{s=0}^l(-1)^s{l \choose s}{2|q|+l \choose |q|+m+s}    \nonumber \\
& & \times   \sum_{t=|q|+m+s}^{N-1-(|q|-m+l-s)}
 (-1)^t e^i_{t,N-1}  \left(\prod_{j \ne i}^N u_j \right) \nonumber \\
 &  \times&  \frac{(N-1-t)!v_i^{N-1-t- | q | +m} }{(N-1-t- | q | +m-l+s)!}  
 \frac{t! u_i^{t- |q| -m} }{(t- |q | -m-s)! }    \, .  \nonumber \\
 & &
 \label{projection_2cf}
\end{eqnarray}
Now the main job is to find an efficient algorithm for determining the polynomials $e^i_{t,N-1}$.
The elementary symmetric polynomials $e_{m,N}$ is defined as 
\begin{eqnarray}
e_{m,N}\left(y_1,...,y_N \right) &=& \left\{ \begin{array}{l}
                                       \sum_{0<i_1<i_2...<i_m \le N}y_{i_1}...y_{i_m}~{\rm for}\,\,m \le N \\
                                       0 \,\,\,\, {\rm otherwise} 
                                     \end{array}  \right.  
                                     \nonumber \\
                                      & &
\end{eqnarray}
The explicit form of the polynomials naturally takes very long time to compute. Davenport and Simon\cite{Simon12} calculated them using
 recursive relations,
\begin{eqnarray}
e_{m,N}(y_1,...,y_N) &=& \frac{1}{m}\sum_{r=1}^{m}(-1)^r p_{r,N}(y_1,...,y_N) \nonumber \\
 & & \times e_{m-r,N}(y_1,...,y_N)  \\ 
  e_{m,N-1}(y_1,...,y_{j\ne i},...,y_N)
 &=& e_{m,N}(y_1,...,y_N) 
   -y_i \nonumber \\
   & \times &  e_{m-1,N-1}(y_1,...,y_{j\ne i},...,y_N) \, , \nonumber \\
   & & 
\end{eqnarray}
where $p_{r,N}(y_1,...,y_N) = \sum_{i=1}^{N}{y_i^r}$ and the restricted polynomial
$e_{m,N-1}(y_1,...,y_{j\ne i},...,y_N) \equiv e^i_{m,N-1}$.
Although the inefficiency occurred for Jain-Kamilla algorithm \cite{JainKamilla97} has been removed in this algorithm, it suffers the numerical issue of precision,
particularly for large $N$, as the value of $y_i^r$ for large $r$ becomes extremely large when $u_i \to 0$. This has been sorted out by calculating the numerical
variables in a higher precision. This, however, extremely slows down the numerical calculation. 
Nonetheless, they were able to find energies for some of the $^2$CF states with negative effective fields.

In case of $^4$CFs with negative effective field, the form of the projected matrix elements will be \cite{Balaram15} 
\begin{eqnarray}
& & \hat{Y}_{|q|,l,m}\left(u_i,v_i\right)J_i^2 \propto 
\sum_{s=0}^l(-1)^s{l \choose s}{2|q|+l \choose |q|+m+s} \nonumber \\
& & \times \sum_{t,t'=0}^{N-1} (-1)^{t+t'} e^i_{t,N-1} e^{i}_{t^{\prime},N-1} \left(\prod_{j \ne i}^N u_j \right)^2  \nonumber \\
& & \times \frac{(2N-2-t-t')! v_i^{2N-2-t-t'- | q | +m }}{(2N-2-t-t'- |q | +m -l +s)!} \nonumber \\
& & \times \frac{(t+t')!u_i^{t+t'-( | q | +m)}}{(t+t'- | q | -m -s)!}
\label{projection_4cf}
\end{eqnarray}
with the summation restricted by the relation
\begin{equation}
 | q | +m +s \leq t+t' \leq 2N-2- | q | +m -l +s \, .
\end{equation}
The numerical evaluation of Eq.(\ref{projection_4cf}) also suffers an issue of precision, even after calculating the variables in the higher precision, 
that tremendously slows down the algorithm. Since we are interested in CFD studies
involving the states consisting of large number of Slater determinants and requiring a huge number of Monte Carlo iterations, the algorithm
is not sufficient for minimizing the time needed for the calculation. We thus suitably modify this algorithm for our purpose.

\subsection{Modified algorithm} 

To get rid of all the denominators in the polynomials $e^i_{t,N-1}$, we redefine the polynomials as 
\begin{equation}
 f_{t}^{i} = \left(\prod_{j\ne i}^{N}u_j\right)e_{t,N-1}^i.
\label{mod_poly} 
\end{equation}
by absorbing $\prod_{j\neq i} u_j$ displayed in Eqs.~(\ref{projection_2cf}) and (\ref{projection_4cf}).
Before calculating the modified polynomial $f_{t}^i$,  we first construct a collection of $(N-1)$ number of sets, defined by $S_t^{u}$ ($1\leq t \leq N-1$) for $N$-th
particle, consisting of $u_k$'s $(k \neq N)$ only. The explicit elements of these sets:
\begin{eqnarray}
&&S_{1}^u = \lbrace{u_1, u_2,\;\dots,\; u_{N-2}, u_{N-1}}\rbrace\nonumber\\
&&S_{2}^u = \lbrace{u_1 u_2, u_1 u_3,\;\dots,\; u_2 u_3, u_2 u_4,\;\dots,\; u_{N-2} u_{N-1} }\rbrace\nonumber\\
&&\cdot \nonumber\\
&&\cdot \nonumber\\
&&\cdot \nonumber\\
&&S_{N-1}^u = \lbrace{u_1 u_2 u_3\; \cdots\; u_{N-2} u_{N-1}}\rbrace  \label{sequence}
\end{eqnarray}
are the all possible products of $t \in [ 1,2,\cdots,(N-1)]$ numbers of $u_k$'s arranged in increasing order of $k$. As a set $S_t^u$ contains $^{N-1}C_t$ number of elements,
the number of elements in $S_t^u$ and $S_{N-1-t}^u$ will be same. We similarly construct the sets $S_t^v$ by replacing $v_k$'s in place of
$u_k$'s in the sets $S_t^u$.
In order to obtain the modified polynomials $f^i_t$, we store all the elements of $S^u_t$ and  $S^v_t$ for $N$-th particle only, 
along with the information about corresponding particle indices of $u_k$'s and $v_k$'s by associating distinct binary numbers, respectively, present in them.


We readily see from the expressions (\ref{mod_poly} and \ref{sequence}) that
\begin{eqnarray}
&&f_{0}^{N} = S_{N-1}^u,\nonumber\\
&&f_{N-1}^{N} = S_{N-1}^v
\end{eqnarray}
are pure functions of $u_k$'s and $v_k$'s, respectively. 
The other modified polynomials of degree $0<t<N-1$ contain both $u_k$'s and $v_k$'s ($k \ne N$). They are  
\begin{equation}
f_{0<t<N-1}^{N} = \sum_{k=1}^{T_t}\lbrace{S_{t}^v}\rbrace_k \, \lbrace{S_{N-1-t}^u}\rbrace_{T_t-k+1} \, ,
\label{poly_N}
\end{equation}
where $\lbrace{S_{t}^v}\rbrace_k$ represents the $k$-th element of the set $S_{t}^v$,  
$\lbrace{S_{N-1-t}^u}\rbrace_{T_t-k+1}$ represents the $(T_t-k+1)$-th element of the set 
$S_{N-1-t}^u$, and $T_t$ is the number of elements in either of the sets $S_t^v$ and $S_{N-1-t}^u$.

Since the term $\lbrace{S_{t}^v}\rbrace_k \, \lbrace{S_{N-1-t}^u}\rbrace_{T_t-k+1}$ in Eq.(\ref{poly_N}) 
contains either the spinor $u_l$ or $v_l$ for any other $l$-th particle, it is possible to    
separate out the terms containing $u_l$ or $v_l$ in $f_t^N$:
\begin{equation}
f_{t}^{N} = L_{t}( u_l ) + R_{t} (v_l) \, ,
\end{equation} 
where $L_{t} (u_l)$ and $ R_{t} (v_l)$ are the sum of the terms containing $u_l$ and $ v_l $
respectively. For any specific particle index $l$,
$R_t(v_l)$ is obtained by collecting terms containing $v_l$; the terms can easily be identified with the above mentioned binary numbers.
The remaining terms automatically constitute $L_t(u_l)$.
The functional form of $L_t (u_l)$ and $R_t (v_l)$ is different for different particles.

Using these $L_{t} (u_l)$ and $ R_{t} (v_l) $, we can easily calculate the polynomials for any particle as
\begin{equation}
f_{t}^{i} =  \left(\frac{u_N}{u_i}\right)L_{t} (u_i) + \left(\frac{v_N}{v_i}\right)R_{t} (v_i) \, .
\end{equation}
Therefore, just by calculating the polynomials for $N$-th particle, we obtain the polynomials for all other particles in a very simple
manner. As we have removed the denominators from the polynomial, this algorithm of calculating different polynomials in recursive method 
does not suffer any  issue of precision in evaluating the wave functions.

Although this modified algorithm has been used only for $^4$CFs with negative effective flux, it is useful for CFs with both negative and positive
effective flux because Eq.~(\ref{mod_poly}) is suitable for any kind of CFs.

\section{PARAFERMIONIC TRIAL STATES}
\label{app_parafermion}

\begin{figure}[h]
\includegraphics[width=7.5cm,angle=0]{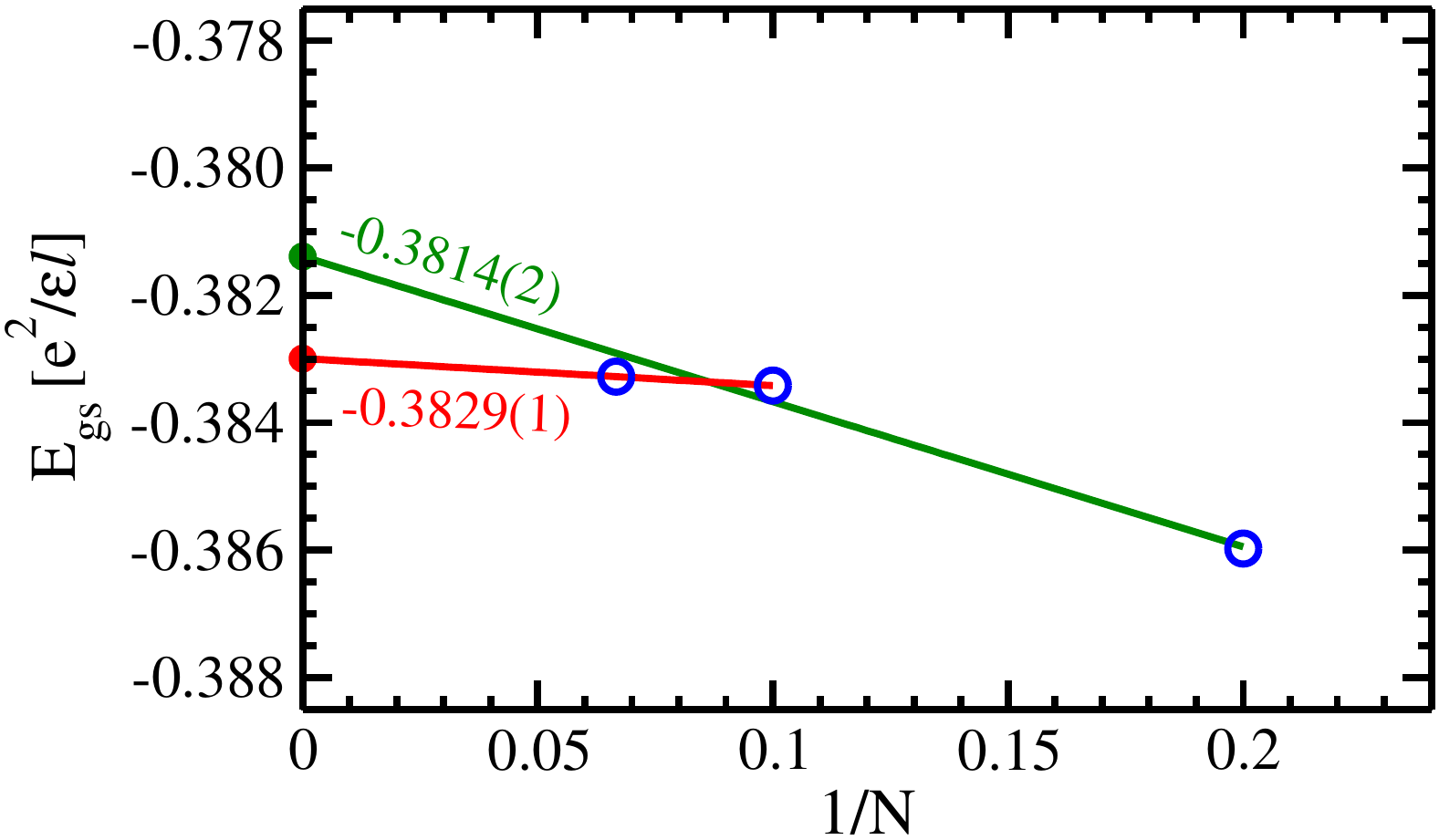}
\caption{(Color online) The ground state interaction energy per particle, $\rm E_{gs}$, obtained from Monte Carlo calculation for the
parafermionic state at $\nu =$ 5/17 for $N=$5, 10, and 15 are plotted against $1/N$ and extrapolated to the thermodynamic limit.
Since the energy for $N=5$ may have large finite size effect, we consider the thermodynamic limit of the energy by linear fitting with $N=10$ and 15
only.}
	\label{gap_parafermion}
\end{figure}

In this appendix, we calculate energy for the previously proposed $Z_5$ parafermion wave function  \cite{Jolicoeur} at 
the filling factor $\nu=5/17$. The  Read-Rezayi (RR) \cite{Read_Rezayi}
parafermion wave functions
are proposed as a generalization of Pfaffian wave function \cite{Moore91} with clustering of composite bosons in $k \geq 2$ clusters.
The corresponding wave function for composite bosons with positive effective flux as a trial wave function \cite{Read_Rezayi,Cappelli,Jolicoeur} for
the ground state of FQHE at
\begin{equation}
\label{Z3_sequence}
\nu = k/(kM+2)
\end{equation}
with the number of clusters $k$ reflecting a state of $Z_k$ parafermions, and the number of vortices attached to each composite boson $M$,
is given by
\begin{equation}
\Psi_{{\rm RR}} = \prod_{i<j}^{N}(u_iv_j-v_iu_j)^M  {\cal S} \left( \prod_{\alpha =1}^k \prod_{i_\alpha < j_\alpha}^{N/k} (u_{i_\alpha}v_{j_\alpha}-u_{j_\alpha}
v_{i_\alpha})^2    \right) .
\label{Z3_state}
\end{equation}
Here ${\cal S}$ is an operator that symmetrizes over all possible clustering of $N$ particles into $k$ sets of equal size.
Equation (\ref{Z3_sequence}) reproduces $\nu = 5/17$, 3/10, and 4/13 for $M=3$ and $k=$ 5, 6, and 8 respectively. 
In spherical geometry,
these parafermionic states have a generalized flux-particle relation $2Q = \nu^{-1}N-5$. 
The general flux-shift $5$ for all these states differ from that (\ref{flux-particle}) in our interacting $^4$CF model . 
With increasing value of $k$, the symmetrization part of the wave function $\Psi_{\rm RR}$ becomes computationally expensive. 
We thus choose $\nu = 5/17$ only for calculating interaction energy using Monte Carlo method. 
Figure \ref{gap_parafermion} shows the energy per electron along with the consideration of the background interaction energy and density
correction factor $\sqrt{2Q\nu/N}$,  for $N=5$, 10, and 15. We find that the thermodynamic limit of the energy of the state $\Psi_{{\rm RR}}$
at $\nu = 5/17$ is $-0.3829(1)\;e^2/\epsilon \ell $ which is reasonably higher than the energy obtained in our CFD calculation. Therefore
the trial wave function $\Psi_{\frac{5}{17}}^{{\rm Unconv}}$ (\ref{Psi_517}) whose overlap with the CFD ground state is very high (Table-\ref{413ovrlp}), 
should be the representative wave function for the filling factor $5/17$.


\begin{thebibliography}{99}
\bibitem{appendix-b} Another wave function for the ground state at $\nu = 5/17$ was proposed using parafermionic physics. We estimate (see Appendix \ref{app_parafermion})
the thermodynamic limit of the interaction
energy for this state and find that it is reasonably higher than the calculated CFD energy here.
\end{thebibliography}
\end{document}